\documentclass[aps,prd,twocolumn]{revtex4}

\newcommand \p{\partial}
\newcommand \br{\nonumber \\ &&}

\def\pmb#1{\setbox0=\hbox{$#1$}%
  \kern-.025em\copy0\kern-\wd0
  \kern.05em\copy0\kern-\wd0
  \kern-.025em\raise.0433em\box0}
\def\parb{\pmb{\partial}}
\def\alt{\mathrel{\hbox{\rlap{\hbox{\lower4pt\hbox{$\sim$}}}\hbox{$<$}}}}

\begin{document}

\title{Well-posedness of the scale-invariant tetrad formulation of the
  vacuum Einstein equations}

\author{David Garfinkle}
\affiliation{Department of Physics, Oakland University, Rochester, MI
48309, USA}
\author{Carsten Gundlach}
\affiliation{School of Mathematics, University of Southampton,
         Southampton SO17 1BJ, UK}

\date{10 Jan 2005, revised 2 May 2005}


\begin{abstract}

We show that with a small modification, the formulation of the
Einstein equations of Uggla et al, which uses tetrad variables
normalised by the expansion, is a mixed symmetric hyperbolic/parabolic
system. Well-posedness of the Cauchy problem follows from a standard theorem.

\end{abstract}


\maketitle


\section{Introduction}

There has been much interest recently in writing the Einstein field 
equations in different ways and analyzing the properties of the 
resulting equations.  Much of the motivation for this comes from 
the search for a form of the equations suitable for numerical 
simulations.  In all these different forms, the Einstein field 
equations form a constrained system.  Since numerical evolutions can 
only approximate the constraints, an analysis of the equations must 
look at the larger solution space where the constraints are not 
satisfied.  In the larger solution space, different formulations 
of the field equations can have
very different properties, in terms both of well-posedness and of 
stability. 

One unusual form of the field equations is due to 
Uggla {\it et al}\cite{Uggla} who use a set of scale invariant 
variables based on the tetrad formalism.  The motivation of \cite{Uggla}
was not to find a system suitable for numerical simulations; but rather
to find a system suitable for the analysis of the properties of singularities,
and in particular of the asymptotic properties of the metric as the
singularity is approached.  For this purpose, well-posedness is not needed:
it is well known that there are forms of the Einstein field equations
for which the initial value formulation is well-posed.  One can then 
take solutions of the field equations (in one of these forms) 
and ask what       
the behavior of the variables of \cite{Uggla} is in these solutions as 
the singularity is approached.
 
Nonetheless, one can examine the equations of \cite{Uggla} to see
whether they have the properties that one would need ({\it e.g.}
well-posedness) for the system to be suitable for numerical simulations.  
Here, the situation does not look hopeful.  A cursory examination of 
the principal part of the equations of \cite{Uggla} shows that it is 
not of any standard type (hyperbolic or parabolic) that would lead one
to expect well-posedness.  The equations of \cite{Uggla} 
were reformulated in \cite{Garfinkle} and then used for a numerical
simulation of the approach to the singularity.  This
reformulation essentially involved using an evolution equation for a
quantity that had previously been given by a constraint.  In the
form used in \cite{Garfinkle} the principal part of the equations 
has the form of a combination of advection and diffusion equations.  
Therefore, one might hope that the system was well-posed.  However, 
no analysis of well-posedness was done in \cite{Garfinkle}.

Here, we perform such an analysis.  We show that the system of
\cite{Garfinkle} is not well-posed as written, but that it can be
made so by a simple modification.  The modification consists of 
adding a multiple of one of the constraint equations to one of 
the evolution eqautions.

In section II we introduce the systems of \cite{Uggla} and 
\cite{Garfinkle}.  Section III contains the analysis of
the well-posedness of the main system, and Section IV analyses the
implied constraint evolution system. Conclusions are given in section V.


\section{Equations}


In \cite{Uggla}  spacetime is described in terms of a set
of coordinates $(t,{x^i})$ and a set of orthonormal vectors (the 
tetrad) $({{\bf e}_0},{{\bf e}_\alpha})$ where both the spatial 
coordinate index $i$ and the spatial triad index $\alpha$ go from 
1 to 3.  The timelike vector  ${\bf e}_0$ of the tetrad 
is chosen to be hypersurface
orthogonal with the relation between tetrad and coordinates chosen
to be of the form ${{\bf e}_0}={N^{-1}}{\partial _t}$ and
${{\bf e}_\alpha}={{e_\alpha }^i}{\partial _i}$.  
The commutators of the tetrad components are decomposed as
follows:
\begin{eqnarray}
\left [ {{\bf e}_0},{{\bf e}_\alpha} \right ] &=& {{\dot u}_\alpha}{{\bf e}_0} -
(H{{\delta _\alpha }^\beta} + {{\sigma _\alpha }^\beta} - 
{{\epsilon _{\alpha \gamma }}^\beta} {\Omega ^\gamma })
{{\bf e}_\beta}  
\\
\left [ {{\bf e}_\alpha} , {{\bf e}_\beta} \right ]   
&=& (2 {a_{[\alpha}}
{{\delta _{\beta ]}}^\gamma}
+ {\epsilon _{\alpha \beta \delta}}{n^{\delta \gamma}})
{{\bf e}_\gamma},
\end{eqnarray}
where $n^{\alpha \beta }$ is symmetric, 
$\sigma ^{\alpha \beta }$ is symmetric and trace free, and 
$\epsilon _{\alpha \beta \gamma}$ is totally antisymmetric
with ${\epsilon _{123}}=1$. 

Scale invariant variables are defined by dividing appropriate quantities
by the expansion $H$.  In particular, scale invariant frame derivatives are
given by
$\{ {\parb _0} , {\parb _\alpha } \} \equiv \{ {{\bf e}_0} , 
{{\bf e}_\alpha } \} / H $
while scale invariant derivatives of the expansion are given by
$q+1 \equiv - {\parb _0} \ln H$ and 
${r_\alpha} \equiv - {\parb _\alpha} \ln H$.
The lapse $N$ is chosen to be equal to $H^{-1}$.  This is equivalent
to the statement that the surfaces of constant time are chosen to 
evolve according to inverse mean curvature flow.  This condition
yields ${\parb _0} ={\partial _t} $ and ${{\dot u}_\alpha } = H {r_\alpha }$.
Other scale invariant variables are given by
\begin{equation}
\{ {{E_\alpha }^i}, {\Sigma _{\alpha \beta }} , {A^\alpha }, 
{N_{\alpha \beta}} \} 
\equiv \{  {{e_\alpha }^i} , 
{\sigma _{\alpha \beta}} 
, {a^\alpha } , {n_{\alpha \beta}} \}  /H
\label{scaleinv}
\end{equation}

In \cite{Uggla} the dynamical system is given by the scale invariant 
variables of equation (\ref{scaleinv}) along with $r_\alpha $.   
The quantity $\Omega ^\alpha$ is not a dynamical variable as it can
be chosen arbitrarily by choosing a rule for the propagation of the
spatial triad ${\bf e}_\alpha$.  In particular, if the spatial triad 
is chosen to be Fermi propagated, then ${\Omega ^\alpha}=0$. 

The quantity $q$ is also not a dynamical variable in \cite{Uggla} as it
is given by
\begin{equation}
 q = {\textstyle \frac 1 3} {\Sigma ^{\alpha \beta}}
{\Sigma _{\alpha \beta}}-{\textstyle \frac 1 3} {\parb _\alpha }
{r^\alpha} + {\textstyle \frac 2 3}{A_\alpha}{r^\alpha}
\label{qdef}
\end{equation} 
where equation (\ref{qdef}) follows from the vacuum Einstein equation.  
Einstein's equation then becomes a set of evolution and constraint equations
for the dynamical variables.  The evolution equations are all first order in 
time, and with the exception of the evolution equation for $r_\alpha$ they
are all first order in space.  The evolution equation for $r_\alpha $ 
can be written as
\begin{equation}
{\partial _t} {r_\alpha} = {\parb _\alpha } q + (q {{\delta _\alpha }^\beta}
- {{\Sigma _\alpha }^\beta} + {{\epsilon _{\alpha \gamma }}^\beta}
{R^\gamma}) {r_\beta}
\label{evr1}
\end{equation}
where ${R^\alpha} \equiv {\Omega ^\alpha}/H$.  However, since $q$ is not
one of the dynamical variables, the quantities $q$ and 
${\parb _\alpha} q$ must be expressed by solving equation (\ref{qdef})
for $q$.  The result is
\begin{eqnarray}
{\partial _t} {r_\alpha} = - {\textstyle \frac 1 3} {\parb _\alpha }
{\parb _\beta}{r^\beta} 
\nonumber
\\
+ {\textstyle \frac 2 3} {\Sigma ^{\beta \gamma}}
{\parb _\alpha} {\Sigma _{\beta \gamma}} + 
(- {{\Sigma _\alpha }^\beta} + {{\epsilon _{\alpha \gamma }}^\beta}
{R^\gamma}) {r_\beta}
\nonumber
\\
+ \left (  {\textstyle \frac 1 3} {\Sigma ^{\beta \gamma}}
{\Sigma _{\beta \gamma}}-{\textstyle \frac 1 3} {\parb _\beta }
{r^\beta} + {\textstyle \frac 2 3}{A_\beta}{r^\beta}
\right ) {r_\alpha}  
\label{evr2}
\end{eqnarray}
The first term on the right hand side of equation (\ref{evr2}) is the one 
with the highest number of derivatives.  It is second order, but not 
positive definite, so the equation is not parabolic.  It does not seem
to be of a type for which results about well-posedness are known and 
does not seem to be the sort of equation for which a numerical 
evolution is likely to be stable.

To address these issues, a slight variation of the system of \cite{Uggla}
was used in \cite{Garfinkle}.  Here, the quantity $q$ was added to the 
list of dynamical variables.  The evolution equation for $r_\alpha$ is
then simply equation (\ref{evr1}) and is first order in space.   Equation
(\ref{qdef}) then becomes a constraint equation.  Since $q$ is now a
dynamical variable, it has an evolution equation which turns out to
be second order in space.  The full set of evolution equations for
the system of \cite{Garfinkle} is     
\begin{eqnarray}
{\partial _t} {{E_\alpha}^i} &=& {{F_\alpha}^\beta}{{E_\beta}^i}
\label{ev1}
\\
{\partial _t} {r_\alpha} &=& {{F_\alpha}^\beta}{r_\beta}+{\parb _\alpha}q
\label{ev2}\\
{\partial _t} {A^\alpha} &=& {{F^\alpha}_\beta}{A^\beta}+ 
{\textstyle \frac 1 2}{\parb  _\beta}{\Sigma ^{\alpha \beta}}
\label{ev3}
\\
\nonumber
{\partial _t} {\Sigma ^{\alpha \beta}} &=& (q-2) {\Sigma ^{\alpha \beta}}
- 2 {{N^{<\alpha}}_\gamma}{N^{\beta > \gamma}} + {{N_\gamma }^\gamma}  
{N^{<\alpha \beta >}} 
\\
&+& {\parb ^{<\alpha}}{r^{\beta >}} 
- {\parb ^{<\alpha}}{A^{\beta >}} 
+ 2{r^{<\alpha }}{A^{\beta >}} 
\\
&+& {\epsilon ^{\gamma \delta < \alpha}}({\parb _\gamma } - 2 {A_\gamma})
{{N^{\beta > }}_\delta}   
\label{ev4}
\\
{\partial _t}{N^{\alpha \beta}} &=& q {N^{\alpha \beta }} + 
2 {{\Sigma ^{(\alpha }}_\delta}{N^{\beta ) \delta }} - 
{\epsilon ^{\gamma \delta (\alpha }}{\parb _\gamma } 
{{\Sigma ^{\beta )}}_\delta}
\label{ev5}
\\
\nonumber
{\partial _t} q &=& \left [ 2 (q-2) + {\textstyle \frac 1 3} 
\left ( 2 {A^\alpha } - {r^\alpha}\right ) {\parb _\alpha}
- {\textstyle \frac 1 3} {\parb ^\alpha}{\parb _\alpha}
\right ] q 
\\
\nonumber
&-& {\textstyle \frac 4 3} {\parb _\alpha}{r^\alpha} + 
{\textstyle \frac 8 3}{A^\alpha}{r_\alpha} + {\textstyle \frac 2 3}
{r_\beta}{\parb _\alpha}{\Sigma ^{\alpha \beta}} 
\\
&-& 2 {\Sigma ^{\alpha \beta}}{W_{\alpha \beta}} 
\label{ev6}
\end{eqnarray}
Here angle brackets denote the symmetric trace-free part, and 
$F_{\alpha \beta }$ and $W_{\alpha \beta} $ are given by
\begin{eqnarray}
{F_{\alpha \beta }} &\equiv & q {\delta _{\alpha \beta}} - {\Sigma _{\alpha \beta}}
\\
\nonumber
{W_{\alpha \beta }} &\equiv & {\textstyle \frac 2 3}{N_{\alpha \gamma}}
{{N_\beta}^\gamma} 
- {\textstyle \frac 1 3} {{N^\gamma }_\gamma}
{N_{\alpha \beta }}
+ {\textstyle \frac 1 3} {\parb _\alpha} 
{A_\beta}
\\
&-& {\textstyle \frac 2 3} {\parb _\alpha} {r_\beta}
- {\textstyle \frac 1 3} 
{{\epsilon ^{\gamma \delta }} _\alpha } \left ( {\parb _\gamma}
- 2 {A_\gamma}\right ) {N_{\beta \delta}}  
\end{eqnarray}
The gauge choice ${\Omega ^\alpha}=0$ is made, which means that the
spatial triad is Fermi propagated.  Note that the highest order term
in equation (\ref{ev6}) is
$ - (1/3) {\parb ^\alpha}{\parb _\alpha} q$ which resembles the 
backward diffusion equation.  Thus, we might expect this equation to be
well behaved for evolution in the negative time direction, which is the
direction appropriate for the approach to the singularity.
   
In addition to the evolution equations, the variables satisfy constraint 
equations as follows:
\begin{eqnarray}
\nonumber
0 &=& {{({{\cal C}_{\rm com}})}^i _{\alpha \beta}} \equiv 
2 ( {\parb _{[\alpha }} - {r_{[\alpha}}-{A_{[\alpha}}){{E_{\beta ]}}^i}
\\
&-& {\epsilon _{\alpha \beta \delta}}{N^{\delta \gamma}}{{E_\gamma }^i}
\label{cn1}
\\
\nonumber
0 &=& {{\cal C}_{\rm G}} \equiv 1 + {\textstyle \frac 1 3} 
(2 {\parb _\alpha} - 2 {r_\alpha} - 3 {A_\alpha}){A^\alpha} -
{\textstyle \frac 1 6}{N_{\alpha \beta}}{N^{\alpha \beta}}
\\
&+&{\textstyle \frac 1 {12}} {{({{N^\alpha}_\alpha})}^2} -
{\textstyle \frac 1 6} {\Sigma _{\alpha \beta}}{\Sigma ^{\alpha \beta}} 
\label{cn2}
\\
\nonumber
0 &=& {{({{\cal C}_{\rm C}})}^\alpha} \equiv {\parb _\beta} 
{\Sigma ^{\alpha \beta}}+ 2 {r^\alpha} - {{\Sigma ^\alpha}_\beta}
{r^\beta} - 3 {A_\beta}{\Sigma ^{\alpha \beta}}
\\
&-&{\epsilon ^{\alpha \beta \gamma}}{N_{\beta \delta}}
{{\Sigma _\gamma}^\delta}   
\label{cn3}
\\
0 &=& {{\cal C}_q} \equiv q - {\textstyle \frac 1 3} {\Sigma ^{\alpha \beta}}
{\Sigma _{\alpha \beta}}+{\textstyle \frac 1 3} {\parb _\alpha }
{r^\alpha} - {\textstyle \frac 2 3}{A_\alpha}{r^\alpha}
\label{cn4}
\\
0 &=& {{({{\cal C}_J})}^\alpha} \equiv ({\parb _\beta} - {r_\beta})
({N^{\alpha \beta}}+{\epsilon ^{\alpha \beta \gamma}}{A_\gamma})
\nonumber
\\
&-& 2 {A_\beta}{N^{\alpha \beta}}
\label{cn5}
\\
0 &=&  {{({{\cal C}_W})}^\alpha}  
\equiv [ {\epsilon ^{\alpha \beta \gamma}}
({\parb _\beta} - {A_\beta}) - {N^{\alpha \gamma}}]{r_\gamma}
\label{cn6} 
\end{eqnarray} 


\section{Analysis of well-posedness}


We now analyze the evolution equations of \cite{Garfinkle} 
(equations (\ref{ev1}-\ref{ev6}) to see whether they are well-posed.
We shall show that with a small modification, the system can be
brought into a form that is mixed parabolic/hyperbolic.  We then use a
textbook theorem to show that the resulting initial value problem
(without boundaries) is well-posed.

The relevant theorems are theorems 4.6.2 and 4.9.1 of \cite{Kreiss} .
Here the first theorem is for linear systems, while the second 
theorem applies the results of the first theorem to nonlinear systems
by considering them as a sequence of linear systems.  For simplicity,
we will present the analysis of a single linear system in the sequence.
That is we assume a guess for the dynamical variables, use that guess
in the coefficients of the evolution equations (\ref{ev1}-\ref{ev6})
and consider the resulting linear problem.  

Theorem 4.6.2 of \cite{Kreiss} asserts the following: 
Assume we have a vector of
variables $u$ and another vector of variables $v$, which obey a
linear system of evolution equations of the form
\begin{eqnarray}
\p_t u &=& D_{11} u + D_{12} v, \\
\p_t v &=& D_{21} u + D_{22} v.
\end{eqnarray}
Here the $D$ are linear spatial derivative operators whose
coefficients can depend on $t$ and $x^i$. $D_{11}$ is
a first-order derivative operator such that $\p_t u=D_{11} u$ is
symmetric hyperbolic. $D_{22}$ is a second-order derivative operator
such that $\p_t v=D_{22}v$ is parabolic. $D_{12}$ and $D_{21}$ are
arbitrary first-order derivative operators. Then the coupled problem
is said to be mixed symmetric hyperbolic/parabolic, and is
well-posed. The theorem assumes periodic boundary conditions, which is
what we are interested in in applications to cosmology. 

Garfinkle's system is first order in time and first order in space,
except for the second derivative in $\p_t q = \dots - 1/3\ \parb_\alpha
\parb^\alpha q$. We therefore try to apply the above theorem to
\begin{equation}
v\equiv q, \qquad u\equiv({E_\alpha}^i,r^\alpha,s^\alpha,
\Sigma^{\alpha\beta},N^{\alpha\beta}), 
\end{equation}
Here to simplify the expressions for the differential operators
we have introduced the notation ${s^\alpha} \equiv
{A^\alpha}-{r^\alpha}$.
Well-posedness depends only on the principal part (here,
terms containing first derivatives) and in examining it we can assume
that the coefficients are constant. In particular, we can treat the
frame derivatives
\begin{equation}
\parb_\alpha \equiv {E_\alpha}^i {\partial\over \partial x^i}
\end{equation}
as if they were partial derivatives, because their commutator is a
lower-order term. In the following, we use the symbol $\p_\alpha$
to stress this.
The principal part of $D_{22}$ is 
\begin{equation}
\label{qdot}
\p_t q\simeq -{1\over 3} \p^\alpha \p_\alpha q.
\end{equation}
Here $\simeq$ stands for ``equal up to lower order terms''. 
It is clear that $\p_t v=D_{22}v$ is parabolic (going backwards in
$t$). Furthermore $D_{11}$, $D_{12}$ and $D_{21}$ are first order. We
only need to check that $\p_t u=D_{11}u$ is symmetric
hyperbolic. 

As in any evolution system subject to constraints, we can change the
level of hyperbolicity of the evolution equations for $u$ by adding
constraints to the right-hand side of the evolution equations. This
does not affect solutions of the evolution equations which also obey
the constraints, but changes the evolution equations off the
constraint surface. The evolution equations are first order and so are
the constraints. Therefore we can add any undifferentiated constraint
to the right-hand side that has the correct number of indices and
symmetries. In the current system, we can add arbitrary multiples of
$(C_C)^\alpha$, $(C_J)^\alpha$ and $(C_W)^\alpha$ to the evolution
equations for $r^\alpha$ and $s^\alpha$. We can also add
$\delta^{\alpha\beta}$ times arbitrary multiples of $C_G$ and $C_q$ to
the evolution equation for $N^{\alpha\beta}$. There are no other
possibilities, and so addition of constraints is determined by eight
free parameters.

The complete calculation is straightforward but tedious. For our
purposes it will be sufficient to add $d(C_C)^\alpha$ to 
$\p_t{s^\alpha}$,
where $d$ is a constant parameter. $d=0$ corresponds to the system of equations
of \cite{Garfinkle} . The principal part of $D_{11}$, making the
symmetrizations explicit, is
\begin{eqnarray}
\label{udotfirst}
\p_t {E_\alpha}^i &\simeq& 0, 
\label{pev1}
\\
\p_t r^\alpha &\simeq& 0, 
\label{pev2}
\\
\p_t s^\alpha &\simeq& \left(d+{1\over 2}\right)\p_\beta \Sigma^{\alpha\beta}, 
\label{pev3}
\\
\p_t \Sigma^{\alpha\beta} &\simeq& 
-{1\over 2}(\p^\alpha s^\beta+\p^\beta s^\alpha) 
+{1\over 3}\delta^{\alpha\beta}\p_\gamma s^\gamma \br
+{1\over 2}(\epsilon^{\gamma\delta\alpha}\p_\gamma {N^\beta}_\delta+\epsilon^{\gamma\delta\beta}\p_\gamma
{N^\alpha}_\delta), 
\label{pev4}
\\
\label{udotlast}
\p_t N^{\alpha\beta} &\simeq&-{1\over 2}(\epsilon^{\gamma\delta\alpha}\p_\gamma{\Sigma^\beta}_\delta
+\epsilon^{\gamma\delta\beta}\p_\gamma{\Sigma^\alpha}_\delta).
\label{pev5}
\end{eqnarray}

In order to show symmetric hyperbolicity, we have to do two separate
things. First we have to find a set of
characteristic variables $\{U\}$ that is complete in the sense of spanning
$\{u\}$. Let $n^\alpha$ be a unit vector with respect to the metric
$\delta _{\alpha\beta}$. Then a characteristic variable 
$U$ with speed $-\lambda$ in
the direction $n^\alpha$ is a linear combination of
$u$ with the property that 
\begin{eqnarray}
\p_t U&=&\lambda n^\alpha \p_\alpha U 
+\hbox{derivatives normal to $n^\alpha$} \br
+\hbox{lower order terms}.
\end{eqnarray}
If all speeds $\lambda$ are real, and the $U$ are complete, the system
is said to be strongly hyperbolic. We introduce a notation where an
index $n$ denotes contraction with $n^\alpha$ or $n_\alpha$, and 
where uppercase Latin indices denote projection with the projector 
\begin{equation}
q_{\alpha\beta}\equiv \delta_{\alpha\beta}-n_\alpha n_\beta.
\end{equation}
In this notation, the principal part of the evolution equations is 
\begin{equation}
\p_t u = P^\alpha \p_\alpha u = (P_n \p_n + P^A \p_A) u.
\end{equation}
The system is strongly hyperbolic if and only if $P_n$ has only real
eigenvalues and is diagonalizable, and if the eigenvalues and the
diagonalizing matrix depend continuously on the direction $n^\alpha$.

Secondly, we have to find an energy density $\epsilon$ that is quadratic in
$\{u\}$, conserved in the sense that there exists a flux $F^\alpha$ so
that 
\begin{equation}
\p_t \epsilon = \p_\alpha F^\alpha +\hbox{lower order terms}
\end{equation}
and that is positive definite in the sense that it is non-negative,
and zero if and only if $u=0$. The system is said to be symmetric
hyperbolic if and only if it is strongly hyperbolic and if it admits a
positive definite conserved energy. 

In order to show strong hyperbolicity, we bring the matrix $P_n$ into
block-diagonal form by considering suitable linear combinations of the
$u$ with respect to $n^\alpha$. 
Any variable whose time derivative is zero and  
whose derivative in the $n^\alpha$ direction does not appear
in the evolution equations
is a characteristic variable with zero speed (called 
a zero mode).  For the purposes of this analysis, each zero mode can be
considered a separate 1 by 1 block of $P_n$ that automatically satisfies 
the conditions needed for strong hyperbolicity.  
From equations (\ref{pev1}-\ref{pev5}) 
it immediately follows that ${E_\alpha}^i$
and $r^\alpha$ are zero modes.

We define $s_n$ as the contraction of
$s_\alpha$ with $n^\alpha$, $s_A$ as its projection with
$q_{\alpha\beta}$, $N_{qq}$ as $N^{\alpha\beta} q_{\alpha\beta}$, and
similarly for other variables. We similarly project the evolution
equations. Using these projections we find that $N_{nn}$ and $N_{qq}$
are zero modes and that for the remaining variables 
$P_n$ is block-diagonal in three blocks,
corresponding to variables which are, respectively, scalars, vectors,
and symmetric tracefree tensors in the space normal to $n^\alpha$. 

The scalar block is given by
\begin{equation}
u^{\rm s}=\left(\begin{array}{c} 
s_n \\ \Sigma_{nn}  \end{array}\right),
\quad
P_n^{\rm s}=\left(\begin{array}{cc}
0 & d+1/2  \\
-2/3 & 0  \\
\end{array}\right).
\end{equation}
Note that because $\Sigma_{\alpha\beta}$ is tracefree (with respect to
$\delta_{\alpha\beta}$), we have
$\Sigma_{nn}+\Sigma_{qq}=0$. Therefore we do not need to consider
$\Sigma_{qq}$ as a separate variable.  $P_n^{\rm s}$ has eigenvalues
$\pm {\sqrt{ -(1+2d)/3}}$, and it is
diagonalizable for $d \ne -1/2$.

The vector block is given by
\begin{equation}
u^{\rm v}=\left(\begin{array}{c} 
s_A  \\ \Sigma_{An} \\ \tilde N_{An} \end{array}\right),
\quad
P_n^{\rm v}=\left(\begin{array}{ccc}
0 & d+1/2 & 0 \\
-1/2 & 0 & 1/2 \\
0 & 1/2 & 0 \end{array}\right).
\end{equation}
Here we have defined the shorthand
\begin{equation}
\tilde N_{An}\equiv \epsilon_{nBA}N^{Bn}.
\end{equation}
$P_n^{\rm v}$ has eigenvalues $(0,\pm\sqrt{-d/2})$, and is
diagonalizable for $d\ne 0$. 

The symmetric tracefree tensor block is given by
\begin{equation}
u^{\rm t}=\left(\begin{array}{c} 
\tilde N_{(AB)} \\ \hat \Sigma_{AB} \end{array}\right),
\quad
P_n^{\rm t}=\left(\begin{array}{cc}
0 & 1 \\
1 & 0 \end{array}\right).
\end{equation}
We have defined the shorthands
\begin{eqnarray}
\hat \Sigma_{AB} &\equiv& \Sigma_{AB}-{1\over 2} q_{AB}\Sigma_{qq}, \\
\tilde N_{AB} &\equiv& \epsilon_{nCA} {N_B}^C.
\end{eqnarray}
Note that it is consistent to remove the $q$-trace from $\Sigma$ as
$\Sigma_{qq}=-\Sigma_{nn}$ appears in the scalar block. Note also
that although the $\delta$-trace of $\Sigma_{\alpha\beta}$ vanishes,
this does not mean that the $q$-trace of the projection $\Sigma_{AB}$
vanishes. Finally, note that $\tilde N_{AB}$ has vanishing $q$-trace
by definition, but is not symmetric. What appears in the tensor block
is its symmetrisation $\tilde N_{(AB)}$. This is consistent because
the antisymmetric part is $\tilde N_{[AB]}=1/2\epsilon_{nAB}N_{qq}$,
and $N_{qq}$ is a zero mode. $P_n^{\rm t}$ has
eigenvalues $\pm 1$, and is diagonalizable for any $d$.

Putting together these results, we see that the $u$ evolution system
is completely ill-posed for $d=0$ (Garfinkle's version of the
equations), but is strongly hyperbolic for $d<-1/2$. The most natural
value is $d=-2$, for which all characteristic speeds are $(0,\pm 1)$.

In order to find the most general $\epsilon$ that is conserved, we
parameterize the most general scalar $\epsilon$ and vector $F^\alpha$
that are quadratic in $u$ without the use of background fields (in
particular $n^\alpha$ must not appear). We then use $\p_t\epsilon
\simeq \p_\alpha F^\alpha$ to determine their coefficients. The result
for $d=-2$ is
\begin{eqnarray}
\epsilon &=& c_0 \left[\left(A_\alpha-r_\alpha\right)^2+{3\over
    2}\left(\Sigma_{\alpha\beta}^2 + N_{\alpha\beta}^2\right)\right] \br
+c_1 r_\alpha^2 
+ c_2 \left({E_\alpha}^i\right)^2
+ c_3 \left({N^\alpha}_\alpha\right)^2 , \\
F^\alpha&=&3c_0\left[\Sigma_{\alpha\beta}(r^\beta-A^\beta)
+\epsilon_{\alpha\beta\gamma}
    N^{\beta\delta}{\Sigma^\gamma}_\delta\right].
\end{eqnarray}
Clearly this is positive definite for $c_0,c_1,c_2>0$ and $c_3\ge
0$. 

We have therefore shown that with the modification $d=-2$, Garfinkle's
version of Uggla et al's equations is mixed symmetric
hyperbolic/parabolic, and from the theorem we have cited the resulting
initial value problem is well-posed.


\section{Constraint propagation}


We should stress that what we have proved is that the Cauchy problem
for the evolution equations (``free evolution'') is well-posed
independently of whether the initial data obey the constraints or
not. This is the well-posedness result required for stability of
numerical free evolution schemes, because numerical error will always
generate small constraint violations even if the initial data obey the
constraints. 

Nevertheless, the constraints are compatible with the evolution
equations in the sense that if we denote the constraints schematically
by a vector $c$, the expression for $\p_t c$ obtained by substituting
the evolution equations is homogeneous in $c$ and $\p_\alpha c$, so
that $c(x,0)=0$ implies $\p_t c(x,0)=0$. In order to guarantee a
well-posed continuum problem solving all Einstein equations (evolution
and constraints) we should check that the constraint evolution is also
well-posed. Together with its closure this guarantees that $c(x,t)=0$
is the {\em unique} solution of $c(x,0)=0$.

It is true in general that if the main system of evolution equations
is strongly hyperbolic and the constraint evolution system closes,
then the constraint evolution system is also strongly hyperbolic
\cite{Reula}. However, there is no such theorem for a mixed
parabolic/hyperbolic problem, and therefore we carry out the analysis
explicitly. The principal part of the constraints is
\begin{eqnarray}
{{({{\cal C}_{\rm com}})}^i _{\alpha \beta}} &\simeq& 
2\p_{[\alpha} E_{\beta]}^i, \\ 
{{\cal C}_{\rm G}} &\simeq& {2\over 3} \p_\alpha A^\alpha, \\ 
{{({{\cal C}_{\rm C}})}^\alpha} &\simeq& \p_\beta
\Sigma^{\alpha\beta}, \\ 
{{\cal C}_q} &\simeq& q + {1\over 3} \p_\alpha r^\alpha, \\ 
{{({{\cal C}_J})}^\alpha} &\simeq& \p_\beta N^{\alpha\beta}
+\epsilon^{\alpha\beta\gamma}\p_\beta A_\gamma, \\
{{({{\cal C}_W})}^\alpha}  &\simeq&  
\epsilon^{\alpha\beta\gamma}\p_\beta r_\gamma.
\end{eqnarray} 
One may wonder at the appearance of the undifferentiated $q$ in
the principal part of
${{\cal C}_q}$, but its presence is required if the principal part of
the constraints is to close under the principal part of the evolution
equations. These are given by (\ref{qdot}) and 
(\ref{udotfirst})-(\ref{udotlast}), with the exception that including
the cross terms in the principal part we must replace
(\ref{pev2}) by
\begin{equation}
\p_t r_\alpha \simeq \p_\alpha q. 
\end{equation}
We find that under the principal part of the evolution equations the
principal part of the constraints evolves as
\begin{eqnarray}
\p_t {{({{\cal C}_{\rm com}})}^i _{\alpha \beta}} &\simeq& 0, \\ 
\p_t {{\cal C}_{\rm G}} &\simeq& {2\over 3} \left(d+{1\over 2}\right)
\p_\alpha {{({{\cal C}_{\rm C}})}^\alpha}, \\
\p_t {{({{\cal C}_{\rm C}})}^\alpha} &\simeq& {1\over 2}
\epsilon^{\alpha\beta\gamma}\p_\beta
\left[({\cal C}_{\rm J})_\gamma
- ({\cal C}_{\rm W})_\gamma\right] \nonumber \\
&-& \p^\alpha\left({\cal C}_{\rm G} -2{\cal C}_{\rm q}\right), \\
\p_t {{\cal C}_q} &\simeq& 0, \\
\p_t {{({{\cal C}_J})}^\alpha} &\simeq& d
\epsilon^{\alpha\beta\gamma}\p_\beta {{({{\cal C}_{\rm C}})}_\gamma}, \\
\p_t {{({{\cal C}_W})}^\alpha}  &\simeq&  0.
\end{eqnarray}
An analysis similar to the one carried out for the main system shows
that the constraint system is strongly hyperbolic for $d<-1/2$, with
the same speeds that arise in the main system. The constraint
evolution system is therefore well-posed if the main system is. With
hindsight the fact that the constraint system does not have a
parabolic part can be explained by the fact that the parabolic part of
the main system is the evolution of the slicing, that is, gauge, which
should not affect the constraint evolution directly.


\section{Conclusions}


We now consider the implications of this result for numerical simulations
of the systems discussed.  For an ill-posed system, unbounded rates of 
growth are achieved in the limit as wavelength goes to zero.
This is what occurs in the scalar sector of our system for $d>-1/2$.  

However, the Einstein equations are well-posed physically in the sense
that they are well-posed in appropriate variables, and so the
instability can only be due to a gauge transformation or to constraint
violation. In the high frequency limit the constraints
(\ref{cn2}-\ref{cn4}) imply that $\partial_\alpha s^\alpha=0$ and
$\partial_\alpha \Sigma^{\alpha\beta}=0$. Therefore, if $n_\alpha$ is
chosen to be the direction of the wave number of the mode, the
components $s_n$ and $\Sigma_{nn}$ (as well as $\Sigma_{An}$) 
vanish if the constraints
vanish. Therefore all scalar modes, and hence all unstable modes for
$d>-1/2$, are constraint violating modes. 

In a numerical simulation, the effective wavelengths
involved are not arbitrarily small; but are instead limited to a size
of order the spatial grid spacing.  The rate of growth of the constraint
violating mode will then be limited by the spatial resolution.  Furthermore,
since the constraints will be enforced (up to numerical truncation error) in
the initial data, the initial amplitude of the unstable mode will be small.

In \cite{Garfinkle} numerical simulations of the $d=0$ system 
(which we have shown here to be ill-posed)
are performed
with a numerical resolution of 50 grid points for each spatial 
dimension.  We have performed numerical simulations of the same initial
data on the same spatial grid for the $d=-0.6$ system (which is
well-posed).  
The results of the
two simulations are effectively identical.  The reason for this is that
the nonlinear terms in the evolution equations tend to damp constraint
violations.  Thus in the $d=0$ evolution the unstable constraint violating
mode (which starts out at very low amplitude) does not have time to grow
to an appreciable amplitude before it is damped by the nonlinear terms.  
(The reason for using $d=-0.6$ rather than $d=-2$ is that in the 
$d=-2$ system lower-order terms
can lead to the growth of constraint violating modes).
However, we have also performed numerical simulations of the the $d=0$
and $d=-2$ systems with a spatial resolution of 1200 gridpoints.  We have
done this by choosing initial data that depends on only one spatial 
coordinate.  Here, the results of the two simulations are very different:
the $d=-2$ simulation yields stable evolution, while the $d=0$ simulation
has an unstable constraint violating mode that grows from a very small
amplitude to an appreciable one. 
Thus, for 
sufficiently good resolution, the modification of our work to the 
evolution equations of \cite{Garfinkle} is needed for a stable numerical
evolution.

\acknowledgments

CG would like to thank Oakland University for hospitality while this
work was carried out.


\end{document}